# VirtuMob : Remote Display Virtualization Solution For Smartphones

M H Soorajprasad[#1], Balapradeep K N[#2], Dr. Antony P J[#3]

[#1] *M.Tech Student, Department of CS&E, KVGCE Sullia, India*
[#2] *Assistant Professor, Department of CS&E, KVGCE Sullia, India*
[#3] *Professor, Department of CS&E, KVGCE Sullia, India*

*Abstract--* **Mobility is an important attribute in today's computing world. Mobile devices, smartphone and tablet PC are becoming an integral part of human life because they are most effective and convenient communication tools. This paper proposes a system to connect and access the desktops of remote computer systems using an android based Smartphone. Virtual Network Computing based architecture is used to develop the proposed system. Through a VirtuMob viewer provided on the user's Smartphone, the user will be able to access and manipulate the desktops of remote computers. Several functionality such as viewing the desktop, mouse operations, keyboard operations, manipulation of documents can be performed from the Smartphone. VirtuMob server should be running on the remote system and it must be attached to a network. VirtuMob Accelerator is used to process the RFB frames of the desktop, perform Encoding of the frames and then relay the frames to the viewer over the internet. Several Encoding techniques are studied and analysed to determine which is best suited for the proposed system.**

## I. INTRODUCTION

Mobile devices, smartphone and tablet pc are increasingly becoming an essential part of human life as the most effective and convenient communication tools not bounded by time and place. The progress of Mobile Computing (MC) becomes an important attribute in the development of IT technology as well as commerce and industry fields [1]. Mobile users collect rich experience of different services from mobile applications like iPhone apps, Google apps etc. which run on the devices and on remote servers over the networks.

Mobility has become an important aspect and rapidly increasing part in today's computing area. High growth has been seen in the development of mobile devices such as, smartphone, PDA, GPS Navigation and laptops with a variety of mobile computing, networking and security technologies. In addition, with the advances in the development of wireless technology like Ad Hoc Network and WiFi, users may be surfing the Internet much easier from before. Thus, the mobile devices have been accepted by more people as the first choice of working and entertainment [2].

Mobile Devices are essentially thin clients whose processing power and storage capability is limited. A thin client computing system composed of a server and a client that communicate over a network using a remote display protocol. The protocol provides graphical contents to be virtualized and served over a network to a client device, and the application logic is executed on the server. Using the remote display protocol, the client transmits user inputs to the server, and the server returns updates of the user interface of the applications to the client. Thin-client computing [3] provides solution to the increasing management complexity and security problems of the present computing systems by increasing continued improvements in network bandwidth, cost, and pervasiveness to return to a more centralized, secure, and easier-to-manage computing strategy.

The Virtual Networking Computing (VNC) system is also a thin client system. VNC also decreases the amount of state maintained at the user's terminal. VNC viewers are thin because they store no state at the endpoint. This systems allows arbitrary disconnection and reconnection of the client with no effect on the session at the server. Since the client can reconnect at a different locations, VNC achieves mobile computing without requiring the user to carry computing hardware [4].

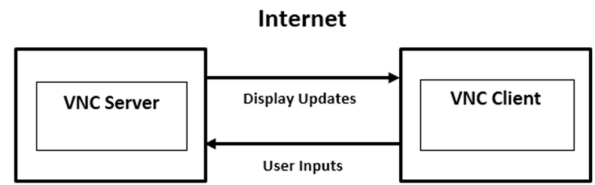

Figure 1. VNC Systems

Remote computing software allows users to remotely access the desktops. The requirement for such remote access is directed by a wide variety of application cases ranging from simple remote access of files and data, to mobile workers who can have access to certain applications only on their work PC, as well as to the remote IT support. The more recent development of virtual desktop infrastructures (VDIs) that almost exclusively based on remote computing software for access has further increased the importance [5].

The quick adoption of smartphones by mobile users has led to users depending on remote computing software on their smartphones to access the PC. Users being able to access their PC from their smartphone has both convenience and productivity implications. Unfortunately, remote computing software available for smartphones are ported on an as-is basis from their PC counterparts and hence do not provide good user experience. The limitations of smartphones in terms of form-factor impose severe overheads on the user, making even simple tasks burdensome.

## II. RELATED WORK

The network computer (NC) objective is to give users access to centralized resources from simple, inexpensive devices. These devices act as clients to a more powerful server machines that are connected to the network and provide applications, data, and storage for user needs. In the virtual network computing (VNC) system [6], server provide not only applications and data but also the desktop environment that can be accessed from any system connected to the internet using a software. When a VNC desktop is accessed, its state and configuration are exactly the same as it was last accessed. Many recent Internet applications have focused on giving users access to resources located anywhere in the world from their home computing systems.

A VNC viewer on one operating system may connect to a VNC server on same or any other operating system. There are clients and servers for many operating systems. Multiple clients may be connected to a VNC server at the same time. Important uses for this technology include remote technical support and accessing files on one's work computer from one's home computer.

The main developments in Virtual Network Computing Systems are as follows.

Tristan Richardson et al [6] proposed a paper on how VNC provides access to home computing environments from anywhere. VNC is a thin client system which is based on a simple display protocol that is platform independent. VNC protocol involves a client and a server and that will operate over any reliable transport such as TCP/IP. The advantages of VNC are 1) Simple and powerful. 2) Appropriate for thin clients. 3) Provides Mobility. 4) Allows a single desktop to be accessed from several places simultaneously. 5) Can be used for wide range of consumer electronic devices.

Kheng-Joo Tan et al. [7] proposed a remote thin client system for the real time multimedia streaming over VNC. Real time transmission is the main challenge with respect to VNC. The thin client systems adopting VNC cannot have a smooth display on multimedia contents through thin clients with the limited data bandwidth. They propose a low complexity Dynamic Image Detection Scheme (DIDS) to split each remote frame to be parts of low motion (like still image) and high motion like video to solve the real time streaming problem. This ensure good video quality in remote thin client systems.

Cynthia Taylor & Joseph Pasquale [8] proposed a way to reduce VNC's poor video performance under high latency conditions. The solution is a very general and simple proxy called a Message Accelerator [MA] that works with VNC to change the effects of network latency.

Cheng-Lin Tsao et al. [9] propose techniques to improve remote computing from smartphones that help deliver PC like experience to users. To improve the performance of remote computing from smartphones they present a solution called Smart-VNC. Smart-VNC incorporates the following four design elements: 1)A new class of application macros called smart-macros that have the principle of traditional raw macros but at the same time possess the robustness of application macros.2) A smartphone friendly interface for the playback of the smart-macros that allows for consistent mixed use of raw input and macros, and allows several acceleration techniques for navigation of the remote computing view on the smartphone.3)Macro parameterization and pre-emption capabilities that grant users to accomplish tasks that are variations of the tasks for which smart macros were created.4) An offline macro recommender that monitors the user activities on a PC and suggests useful smart-macros that depicts tasks frequently performed by the user. The SmartVNC solution allows users to create expandable and robust smart-macros on the PC for any application, and request them from the smartphone using a simple interface overlaid on the default VNC client.

## III. PROPOSED SYSTEM

The major problems with respect to VNC and MC are as follows.

1) VNC are not designed for remote access from a smartphone. Remote access from a smartphone is comparatively more difficult because of device constraints like a small screen size, and the lack of a physical mouse/keyboard.
2) Real time transmission to the thin clients is the important challenge. The thin client systems incorporating VNC cannot have a smooth display on multimedia contents under the limited data bandwidth.
3) Thin client systems may suffer from poor performance due to network latency. Network latency affects every message sent between the client and server.
4) Changing wireless network conditions, limited battery lifetime and latency for interaction are the challenges for remote display in mobile devices.

Hence the proposed system should be able to overcome the above problems and should have the following features. 1) It should be able to send heavy data content through a low bandwidth network. 2) The content displayed at the viewer should be of good quality. 3) Real-time processing of the user interactions. 4) The complexity of the system should not exceed the features of the mobile client in terms of CPU, memory and battery.

The architecture design of proposed system is as shown in the Fig.2. The proposed system VirtuMob consists of a server and a client that communicate over a network using a remote display protocol. The graphical displays on the server side are to be virtualized and provided across a network to a client device. All the applications are executed on the server. The client transmits user interactions to the server, and the server provides screen updates of the user interface of the applications to the client. The thin-client approach provides several significant advantages over traditional desktop computing. Clients can be essentially stateless devices that do not need to be backed up or restored and do not store any sensitive data that can be lost or stolen. Mobile users can access the server from any client and obtain the same persistent, personalized computing environment. The data flow diagram of the proposed system is shown in the figure 3.

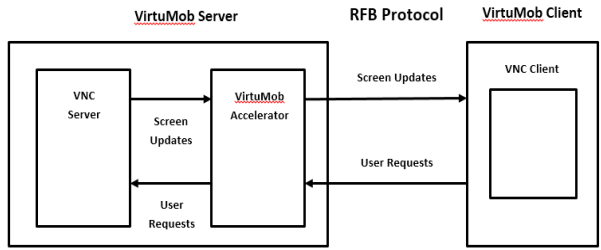

Fig 2. System Architecture

The VirtuMob System consists of VirtuMob Server, VirtuMob Accelerator. The Client request for connection establishment to the server. The Server checks for the IP address and port information and connects to the client. VirtuMob Server then starts relaying the frames to the Accelerator. The Accelerator encodes and compresses the frames and relays it to the client device. Client device relays the user inputs to the Server. These interactions include Key and Mouse events, Modification of documents and request for termination of the connection.

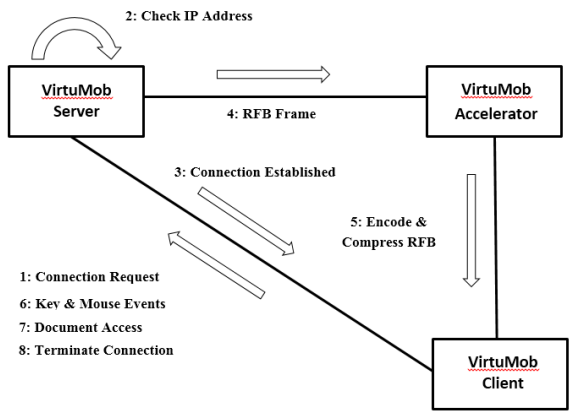

Fig 3. Data Flow Diagram

## IV. ENCODING TECHNIQUES

The VNC protocol functioning consists of accepting the request from the client about a specific onscreen pixel and then server responding in the form of the update. This update consists of an encoding the difference between the moment of the request and the last time the client requested data about this rectangle. Sending of complete raw information will lead to the high consumption of the bandwidth with the consequent delay in the process. Encoding refers to the method in which a rectangle of pixel data will be provided to the client. Every pixel data is prefixed by a header .The header provides info about the position of the rectangle on the screen, the height and width of the rectangle and type of encoding. The encoding type specifies the way in which encoding of the pixel data is done. The data itself then sent using the defined encoding. These encoding are used to determine the ways to provide the graphical information. When a client wants to establish communication with server both side must negotiate the encoding type to be used. If the client requires a different encoding, the server will specify the next encoding available.

Following are the types of VNC encoding that can be used:

*1.) RAW*

*RAW* is the simplest form of encoding. Here server sends to the client all graphical pixels in the form of width*height pixel values (where width and height are the width and height of the rectangle). This encoding type method must be supported by clients. The processing time used is low and the performance is high when the server and the client are on the same device. If the client is a remote device the performance is reduced due to the transfer of large amounts of data. This encoding is designed ideally for the low performing devices.

*2.) RRE*

RRE stands for Rise-and-Run-length-Encoding which consists of grouping the consecutive identical pixels in order to send only the information of one pixel and the total number of replications of that pixel. The basic idea behind RRE is the partitioning of rectangle of pixel data into sub regions such that each consist of the single pixel value and union of it comprises the original rectangular region. It is the most effective method when large blocks of the same color exist. There is a variation of the protocol which uses a maximum of 255x255 pixels to reduce the size of the packages. RRE rectangles provided to the client are in the form which are easily rendered instantly and efficiently by any graphics engines.

*3.) CopyRect encoding* (*copy rectangle*)

This encoding is a very simple and efficient encoding. This is used when the client already has the same pixel data in its frame buffer. The encoding consists of an X,Y coordinate and provides a position in the frame buffer from which the client can copy the pixel data. This can be used in many situations like when the user moves a window across the screen, and when the window are scrolled. Another major use is for optimizing drawing of text or other repeating patterns.

*4.) Hextile*

This encoding divides the pixels in the 16*16 tiles and allowing the dimensions to be specified in 4 bits each and 16 bits in total. The rectangle is split into tiles beginning at the top left going in left-to-right, top-to-bottom order. This encoded data contents of the tiles follow one another in the predefined order. If the width of the whole rectangle is not an exact multiple of 16 then the width of the last tile in each row will be smaller than that of previous tiles. Each of this tile is either encoded as raw pixel data, or as a variation on RRE. And each tile has a background pixel value. No need to specify for any given tile if it is same as background of the previous tile.

*5.) Zlib Encoding*

Zlib Encoding uses a method to compress the information in order to reduce the size of the package as much as possible. The main drawback is that it requires a great amount of processing.

## V. EXPERIMENT

The different encodings were analyzed and a test scenario of workload for the system has been designed. The objective of this experiment is to determine which of the encodings under analysis is the most suitable for use within the proposed architecture. The different encodings used for the study are RAW, Hextile and Zlib. A scenario of activities to be performed is developed to analyze the performance of the encodings. The list of activities is as follows:

- The mobile device is in the HOME at the beginning.
- Opening the browser.
- Waiting for 3 seconds.
- Opening the music player.
- Waiting for 3 seconds.
- Returning back to the HOME
- End of the benchmark.

This scenario generates a display of around 10 seconds in which the device screen is constantly changing and updating. The system is provided a limited situation in terms of visualization of the device. Since a manual execution would not provide completely reliable data, an automatic test case is used to achieve the same set of execution for every encoding test.

## VI. PERFORMANCE ANALYSIS

The different encoding techniques are analyzed and the results of the analysis are shown in the Table 1. Hextile encoding method provide a high number of updates to the client. But Zlib sends the greatest amount of information with the lowest size of data exchanged. Hextile protocol also exchanges a small amount of data due to its compression mechanism, but the quality of the graphical information is significantly lower. Also, as Fig. 4 shows, Hextile has the problem of exchanging almost more than the the data exchanged by Zlib, and thereby occupying the bandwidth. Hextile pretty much takes up a lot of the bandwidth as it transmits large amounts of data.

TABLE 1
RESULTS OF THE ANALYSIS

|  | **RAW** | **Hextile** | **Zlib** |
|---|---|---|---|
| Updates | 8 | 20 | 68 |
| Updates/second | 0.32 | 0.82 | 1.65 |
| Rectangles received | 8 | 22 | 808 |
| Data captured (MB) | 10.10 | 26.53 | 91.70 |
| Data compressed (MB) | 10.10 | 5.64 | 8.90 |
| Compression ratio | 1 | 4.70 | 10.30 |

The VirtuMob system sends lower amounts of data due to the slower connection speed. The number of updates is less and hence less is captured. In this scenario, the compression provided by the encoding allows the system to send more updates regardless of the speed problem, as shown in Fig. 4. The compression ratio of Zlib and Hextile encoding converts this encoding to the more suitable option so that it can be used in this type of scenario. As a result of the study, it is clear that the Zlib encoding offers the best performance in any of the scenarios. It provides a smooth display that will offer a correct remote visualization to the user. Also it takes up a small amount of data through the network.

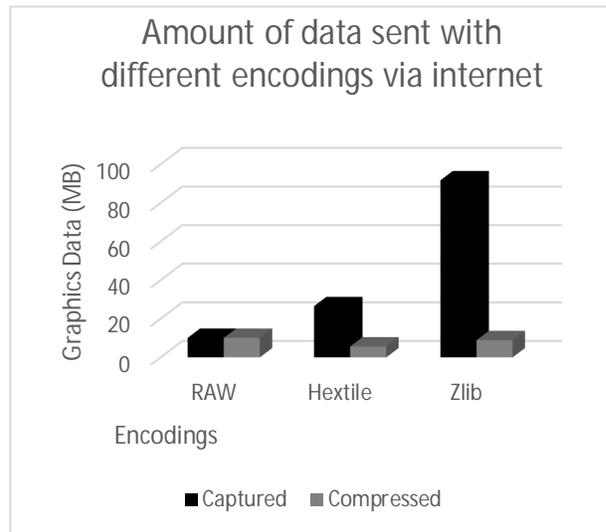

Figure 4. Amount of data sent with different encodings.

## VII. CONCLUSION

The VirtuMob System provides user the capability to use their smartphones to connect and access a desktop system located at a remote place over the internet. The graphical displays are virtualized and provided across a network to a smartphone. The user inputs to the server are provided on the smartphone, and the server returns screen updates of the user interface of the applications to the client. The VirtuMob system provides good quality of the displayed content. It can also transmit heavy content through a bandwidth constrained network. The user interactions are processed in real-time which provides smooth display and interaction. Also the different encodings are analysed that are used to encode the frames. From the experimental study it is found that the zlib encoding technique is suitable for our system. In summary VirtuMob system is an efficient remote display virtualization solution for a smartphone.